\documentclass[12pt]{article}
\usepackage{psfig}
\newcommand{\sun}{\odot}
\renewcommand{\baselinestretch}{1.5}
\begin{document}
\begin{center}
{\Large \bf HI Observations of Flat Galaxies}
\bigskip

{\large S. N. Mitronova$^1$, W. K. Huchtmeier$^2$, I. D. Karachentsev$^1$,
V. E. Karachentseva$^3$, and Yu. N. Kudrya$^3$}

\bigskip

{\em $^1$ Special Astrophysical Observatory, Russian Academy of Sciences,
Nizhnii Arkhyz, Karachai-Cherkessian Republic, 357147 Russia}

{\em $^2$ Max Planck Institut f\"ur Radioastronomie, Auf dem H\"{u}gel 69,
D-53121 Bonn, Germany}

{\em $^3$ Astronomical Observatory,
Kiev National University, Observatorna ul., Kiev, 304053 Ukraine}
\end{center}

\begin{abstract}
We present the HI observations of 94 flat spiral
galaxies from RFGC (the Revised Flat Galaxy Catalog) and 14
galaxies from 2MFGC (the 2MASS selected Flat Galaxy Catalog)
 performed with the 100-m radio telescope in Effelsberg (Germany).
HI fluxes, heliocentric radial velocities, and HI line widths are
 given for 65 detected galaxies. We present a mosaic of HI profiles.
 We calculated some of the global parameters of the galaxies and
analyzed the linear correlations between them. The ratios of the
total (indicative) masses of the galaxies to their luminosities
lie within the range 0.4 with a mean of 3.8 ($M_{\sun}/L_{\sun}$), and
the mean mass fraction of neutral hydrogen is 13\%. Upper limits
are given for the radio fluxes from 43 undetected galaxies.

\end{abstract}

\hspace{5cm}

\newpage

\bigskip

{\large  INTRODUCTION}

\bigskip

One of the central problems in extragalactic astronomy is to study
the collective motions of galaxies by analyzing their peculiar
velocities $V_{pec} = V_{3K} - Hr$ . Here, $V_{3K}$ is the measured radial
velocity of the galaxy reduced to the frame of the $3K$ microwave
background radiation (Kogut et al. 1993), and $Hr$ is the distance
to the galaxy (in km s$^{-1}$ ) determined independently of its radial
velocity. The main method of determining $Hr$ for spiral galaxies is
based on the Tully--Fisher (1977) relation between the absolute
parameters of the galaxy (its luminosity and linear diameter) and
the width of the 21-cm HI line. Karachentsev (1989) showed that
the late-type edge-on spiral galaxies are appropriate objects for
investigating the large-scale streams for several reasons: (1)
the flat, disklike galaxies have a simple structure; (2) applying
the simple selection criterion based on the apparent axial ratio
$a/b > 7$ to them yields a morphologically homogeneous sample; (3)
the detection probability of such galaxies in the 21-cm line is
very high; and (4) since the flat galaxies are located mostly
outside groups and clusters, their structure remains undistorted
and they are not affected by large virial motions. To compile our
catalog of flat edge-on galaxies, we conducted an all-sky survey
using blue and red
POSS-I and ESO/SERC maps. FGC (the Flat Galaxy Catalog; Karachentsev
et al. 1993) includes galaxies with apparent axial ratios $a/b > 7$ and angular
 diameters $a > 0.6^{\prime}$. The updated and supplemented version of this catalog,
RFGC (the Revised Flat Galaxy Catalog; Karachentsev et al. 1999a), contains
4236 galaxies and is one of the best samples for studying the large-scale
cosmic streams of galaxies (Karachentsev et al. 2000a; Kudrya et al. 2003),
since it is highly homogeneous and has the required completeness (Kudrya
et al. 1996, 1997; Feldman et al. 2003).

Extensive high-accuracy
measurements of the radial velocities ($V_h$ ) and HI-line widths ($W$ )
are needed to use the Tully--Fisher (TF) relation. For this purpose,
flat galaxies were observed with large radio telescopes: the 305-m radio
telescope in Arecibo (Giovanelli et al. 1997) and the 100-m radio
telescope in Effelsberg (Huchtmeier et al. 2005). Makarov et al.
(2001) measured the rotation curves of 300 flat galaxies with the
6-m telescope at the Special Astrophysical Observatory of the Russian
Academy of Sciences. Based on original observations and published data,
Karachentsev et al. (2000b) compiled a list of peculiar velocities for
1327 RFGC galaxies. The published homogeneous and complete near infrared
all-sky survey , 2MASS (the Two Micron All-Sky Survey; Skrutskie et al.
1997), opened up
new opportunities for studying the collective motions of galaxies based
on the near-infrared TF relation (Karachentsev et al. 2002; Kudrya et al.
2003).
2MASS J-, H-, and K-band photometry is available for 71\% of the RFGC
galaxies, but the radial velocities and HI-line widths were reliably
measured
only for 25\% of them. A program of HI observations of RFGC galaxies
with the 100-m radio telescope in Effelsberg was initiated in 2001.
Huchtmeier et al. (2005) published the first results of observations
of 268 galaxies.

Analyzing the properties of RFGC galaxies in the near
infrared (2MASS) and comparing them with optical data allowed the
selection criteria to be developed for compiling a new catalog of
flat galaxies based on 2MASS. The 2MFGC all-sky catalog of disklike
galaxies (Mitronova et al. 2004) contains IR photometry and LEDA and
NED identifications for 18020 objects with XSC 2MASS axial ratios $a/b\geq 3$.

In this paper, we present the observations of 94 RFGC galaxies and the
test observations of 14 2MFGC galaxies performed with the 100-m radio
telescope in the fall of 2004.

\bigskip

{\large  HI OBSERVATIONS}

\bigskip

 The HI observations of
RFGC galaxies have been performed with the 100-m radio telescope in
Effelsberg (Germany) since October 2001. Using the optical and infrared
parameters of flat galaxies, we estimated their expected radial
velocities and selected objects in the working range of the telescope
$V_h\leq$ 9500 km s$^{-1}$ for our observations. We also included in our
observational
application galaxies for which
the radial velocities were known from optical observations with an accuracy
lower than 30 km s$^{-1}$ and no HI line width measurements were available.
We
also included in our list galaxies for which it was necessary to improve
$V_h$ and (or) $W$, since they deviated by more than 3$\sigma$ from the
regression
line in the TF relation or the magnitudes of their peculiar velocities
were larger than 3000 km s$^{-1}$ . The spectroscopic observations with the
100-m radio telescope were performed in full energy mode (ON-OFF),
combining the measurements in the source's field with those in the
comparison field whose location was earlier than the galaxy's location
by 5 min in right ascension. The beam FWHM
of the telescope at 21 cm is 9.3$^{\prime}$. The temperature of the system
consisting
of two receivers is $30K$. The 1024-channel autocorrelator was divided into
four 256-channel bands shifted in frequency by 11 MHz. To cover the entire
velocity range from 250 to 9050 km s$^{-1}$ , we used 12.5-MHz-wide bands. In
this case, the resulting resolution per channel is 10.4 km s$^{-1}$. For
galaxies with known radial velocities, we used 6.25-MHz-wide bands,
which provided a resolution of 5.2 km s$^{-1}$ per channel. We used weighted
(or equivalent) smoothing for most galaxies to improve the signal-to-noise
ratio. From one to five scans were made for each galaxy; the total
accumulation time was about $2^h$ per galaxy.

\bigskip

{\large  RESULTS OF THE OBSERVATIONS}

\bigskip

Figures 1 and 2 show the HI profiles for the detected RFGC and 2MFGC
galaxies. No HI line profiles are shown for two RFGC galaxies (2553
and 4160) with complex spectra.

Table 1 contains the results of our
radio observations of 94 RFGC galaxies as well as cataloged (optical)
data. The columns of this table give the following: (1) the RFGC galaxy
number; (2) the J2000.0 equatorial coordinates; (3) the major and minor
diameters ($a\times b$) in arcmin corresponding to the $B$ band 25 mag arcsec$^{-2}$
isophote; (4) the total B -band magnitudes; (5) the morphological type
in the Hubble system, where Sb = 3, Sc = 5, and Sd = 7; (6) the HI-line
flux in Jy km s$^{-1}$ corrected for the ratio of the galaxy's angular
diameter
to the telescope's aperture; (7) the maximum flux and the rms noise error
in mJy; (8) the mean heliocentric radial velocity and its error in km s$^{-1}$;
(9--11) the HI line widths at 50\%, 25\%, and 20\% of the peak flux level
in km s$^{-1}$ . The colons (:) in the table mark unreliable measurements with
low signal-to-noise ratios and a complex structure of the HI profiles.

Table 2 gives data for 2MFGC galaxies. The content of its columns
corresponds to Table 1 with the following differences: (1) the galaxy
names in known catalogs taken from the NED electronic database; (3) the
major and minor diameters ($a\times b$) in arcmin taken from the LEDA electronic
database; and (4) the total $B$ -band magnitudes taken from the LEDA database.

For 40 RFGC and three 2MFGC galaxies, we failed to find the HI line in the
range of radial velocities under study. Tables 1 and 2 give only the rms
noise errors for these galaxies. The undetected galaxies probably lie
outside the working spectral range, although it may well be that the HI
fluxes from some of the nearby early-type galaxies could be below the
instrumental detection threshold.

\bigskip

{\large GLOBAL OPTICAL AND HI PARAMETERS OF GALAXIES}

\bigskip

 We calculated the global characteristics of 54 detected RFGC galaxies
using the radio observations and optical parameters from Table 1. As the
distance indicator, we used the radial velocity of the galaxy reduced to
the centroid of the Local Group,

\begin{equation}
 V_{LG} = V_h + V_a[cos b\cdot cos (b_a)\cdot cos (l-l_a) + sin b\cdot sin (b_a)],
\end{equation}

where $l$ and $b$ are the Calactic coordinates, and $V_h$ is the measured
heliocentric radial velocity; the apex parameters,
$V_a$ = 316 km c$^{-1}$, $l_a = 93^{\circ}$, $b_a =-4^{\circ}$,
were taken from the paper by Karachentsev and Makarov (1996).
We calculated the HI mass as
\begin{equation}
    \log (M_{HI}/M_{\sun}) = \log F + 2\log (V_{LG}/H_{0}) + 5.37,
\end{equation}
where the Hubble constant was assumed to be $H_0$ = 75 km s$^{-1}$ Mpc$^{-1}$,
 and $F$
is the detected 21-cm flux in Jy km s$^{-1}$ corrected for the ratio of the
galaxy's angular diameter to the telescope's aperture. We calculated the
indicative mass of the galaxy within the standard optical diameter from
the relation
\begin{equation}
 \log(M_{25}/M_{\sun}) = 2\log(W_{c})+\log(a_{c})+\log(V_{LG}/H_{0}) +3.92,
\end{equation}
where $W_{c}$ is the hydrogen line width corrected for turbulence (Tully and
Fou\'que 1985) and relativistic broadening, $W_c = W_{50} /(1 + V_h /c)$.
The angular
diameter was reduced to the 25 mag arcsec$^{-2}$ isophote (Kudrya et al.
1997)
 and corrected for the Galactic absorption and the galaxy's tilt
($\log a_{c} = \log a + 0.09A_{B} - 0.2\log (a/b)$). To calculate the
luminosity,
we use the relation
\begin{equation}
  \log (L/L_{\sun}) = 2 \log(V_{LG}/H_{0}) - 0.4(B_t - \Delta B) + 12.16,
\end{equation}
where the total magnitude $B_t$ was corrected for the absorption $A_B$
(Schlegel et al. 1998) and the galaxy's tilt $\Delta B = A_{B} + 1.2\log (a/b)$.

Table 3 gives the calculated global parameters. Its columns contain
the following: (1) the RFGC galaxy number; (2) the radial velocity
$V_{LG}$ reduced to the centroid of the Local Group; (3) the HI line width
$W_c$ at 50\% of the peak flux level corrected for turbulence and
relativistic broadening; (4) the logarithm of the HI mass,
$log(M_{HI}/M_{\sun})$;
(5) the logarithm of the galaxy's indicative mass, $log(M_{25}/M_{\sun})$;
(6) the logarithm of the galaxy's total luminosity, $log(L/L_{\sun})$;
(7) the HI mass-to-luminosity ratio,  $M_{HI}/L$;
(8) the galaxy's total mass-to-luminosity ratio, $M_{25}/L$.
(the values in columns 4--8 are in solar units); and (9)
the mass fraction of neutral hydrogen, $M_{HI}/M_{25}$.

 According
to Table 3, the mean depth of the sample is 4823 km s$^{-1}$ , which
corresponds to 64 Mpc. The total mass-to-luminosity ratios of the
galaxies lie
within the range 0.4--8.2 with a mean of 3.8, and the mass fraction
of neutral hydrogen is 13\%.

Karachentsev et al. (1999b) performed a
statistical analysis of the global parameters for 587 flat spiral
galaxies using the observations performed by Giovanelli et al. (1997)
with the Arecibo radio telescope. Huchtmeier et al. (2005) presented
the statistics of global parameters for 121 detected RFGC galaxies.

Table 4 gives the statistical characteristics of the distribution of
50 flat galaxies in integrated parameters. Comparison of the mean
values of all parameters with those for the sample of 121 flat galaxies
shows that they are almost equal.

In Figs 3 and 4, the logarithm of the
HI mass-to-luminosity ratio and the logarithm of the indicative
mass-to-luminosity ratio are plotted against radial velocity $V_{LG}$
for flat galaxies. Since the radial velocity is a distance indicator,
it can be noted that $M_{HI}/L$
and $M_{25}/L$ decrease with distance only slightly; i.e., there is virtually
no distance selection effect. Arbitrarily oriented spiral galaxies are known
to show no variation of $M/L$ with distance either (Roberts and Haynes 1994).

Figure 5 shows the distribution of galaxies in morphological type. All of
the 94 RFGC galaxies under study are shown on the histogram, and the types
of 54 detected galaxies of them are marked by double hatching. Most of the
objects belong to the morphological types $Sbc - Sd$, where the detection
level is higher.

The data in Table 5 agree with the conclusions of our
previous paper (Huchtmeier et al. 2005) that the mean values of the global
parameters under consideration do not change within the error limits for
galaxies of various morphological types. This conclusion is consistent with
the HI observations of edge-on spiral galaxies (Karachentsev et al. 1999b)
and arbitrarily oriented spirals of various types (Haynes and Giovanelli
1984) from the catalog of isolated galaxies by Karachentseva (1973).

\bigskip

{\large BIVARIATE DISTRIBUTIONS OF GLOBAL PARAMETERS}

\bigskip

The correlations between global parameters lead us to important
conclusions about the differences between the properties of the disk
structures in giant and dwarf galaxies. The observed differences can
characterize the different formation and equilibrium conditions for
the gaseous disks and the inequality of the star formation rates and
intensities in them. It is also assumed that the contribution of dark
matter depends on the linear sizes of galaxies.

Figures 6 and 7 show
the bivariate distributions of global parameters for 50 detected RFGC
galaxies, and Table 6 lists the parameters of the linear regression
$y = kx + c$ for various relations. The logarithms of the various global
parameters of flat galaxies are the variables $x$ and $y$ . Columns 4 and
5 give the correlation coefficients $\rho(x, y$) and the standard deviations
$\sigma (y)$; columns 6 and 7 contain the regression parameters $k$ and
$c$ and their
standard errors, respectively. Figure 6a and the first row of Table 6 show
the relation between the logarithm of the HI mass (in
$M_{\sun}$) ) and the logarithm of the galaxy's linear diameter $A_{25}$
(in kpc) calculated from the angular diameter ac using the formula
\begin{equation}
 A_{25} = 0.29\cdot a_c \cdot V_{LG}/H_0.
\end{equation}
The linear relation suggests that the HI
density in the disks of spiral galaxies is virtually constant and does
not depend on the linear size. The conditions for star formation in them
appear to have been approximately identical in them. The statistically
insignificant decrease in log $(M_{HI}/A_{25}^2)$ toward the massive
galaxies
(row 10 in Table 6) may be due to the internal absorption of the HI
flux in the disks of the largest galaxies. Other parameters, the indicative
mass and the total luminosity (see rows 2 and 3 in Table 6 and Fig. 6b),
also correlate well with the linear diameter.

The dependences of the HI
mass log $M_{HI}$ and the indicative mass log $M_{25}$ on the luminosity
log $L$ show
a systematic decrease in $M/L$ from dwarf galaxies to giant spirals
(see Figs. 6c and 6d and rows 4 and 5 in Table 6). This is to be expected
if the relative abundance of the dark matter in giant galaxies is lower
than that in dwarf galaxies.

Since the width of the 21-cm HI line does
not depend on the distance to the galaxy, taken as the argument for the
bivariate distributions of global parameters, it allows the morphology
and luminosity selection effects to be reduced. Figure 6e (row 6 in Table 6)
shows the linear relation between the galaxy's size log $A_{25}$ (in kpc) and
the amplitude of its internal rotation determined as the width of the 21-cm
line. The slope of the linear regression in this Tully--Fisher (1977)
relation is 1.28 with a standard deviation of 0.15. The correlations
between log$(M_{25}/L)$ and log$(M_{HI}/L)$ and the 21-cm line width are also
well defined; their parameters are listed in rows 7 and 8 of Table 6.
We see from Fig. 6f (row 9 in Table 6) that the relative hydrogen
abundance increases from giant galaxies to dwarf galaxies in the same
way as for arbitrarily oriented spiral galaxies whose luminosities are
not affected strongly by internal absorption (Huchtmeier and Richter
1988, Stavely-Smith and Davies 1988).

The relation between the logarithm
of the HI mass, log $M_{HI}$, and the specific angular momentum log$(A_{25} W_{50}$ )
(see Fig. 7 and row 11 in Table 6) has a high correlation coefficient;
the slope of the linear regression is 1.19. Zasov (1974) pointed out
that such a relation must hold in the gas-rich disks of spiral galaxies
with active star formation. Karachentsev et al. (1999b) corroborated the
linear relation between log $M_{HI}$ and log$(A_{25} W_{50}$ ) for flat
galaxies and
Zasov's conclusion that the gaseous disks of galaxies must be near the
gravitational stability boundary irrespective of the linear sizes of
the galaxies.

\bigskip

{\large  ACKNOWLEDGMENTS}

\bigskip

 We are grateful to Max Plank Institut f\"ur Radioastronomie (Effelsberg,
Germany) for the observational data obtained with the 100-m radio telescope.
This work was supported by the Russian Foundation for Basic Research and
Deutsche Forschungsgemeinschaft (DFG) (grant no. 02--02--04012).

\bigskip

{\large  REFERENCES}

\bigskip

1. H. Feldman, R. Yuszhkievicz, M. Davis, et al., Astrophys. J. 596, L131
(2003).

2. R. Giovanelli, E. Avera, and I. D. Karachentsev, Astron. J. 114,
122 (1997).

3. M. P. Haynes and R. Giovanelli, Astron. J. 89, 758 (1984).

4. W. K. Huchtmeier, Yu. N. Kudrya, I. D. Karachentsev, et al.,
Astron. Astrophys. (2005) (in press).

5. W. K. Huchtmeier and O. G. Richter, Astron. Astrophys. 203, 237 (1988).

6. I. D. Karachentsev, Astron. J. 97, 1566 (1989).

7. I. D. Karachentsev et al., Astron. Astrophys. (2004) (in press).

8. I. D. Karachentsev, V. E. Karachentseva, Yu. N. Kudrya,
et al., Bull. Spec. Astron. Obs. Russ. Acad. Sci. 47, 5 (1999).

9. I. D. Karachentsev, V. E. Karachentseva, and Yu. N. Kudrya,
Pis'ma Astron. Zh. 25, 3 (1999b) [Astron. Lett. 25, 1 (1999b)].

10. I. D. Karachentsev, V. E. Karachentseva, Yu. N. Kudrya, et al.,
Astron. Zh. 77, 175 (2000a) [Astron. Rep. 33, 150 (2000a)].

11. I. D. Karachentsev, V. E. Karachentseva, Yu. N. Kudrya, et al.,
Bull. Spec. Astron. Obs. Russ. Acad. Sci. 50, 5 (2000b).

12. I. D. Karachentsev, V. E. Karachentseva, and S. L. Parnovskii,
Astron. Nachr. 314, 97 (1993).

13. I. D. Karachentsev and D. I. Makarov, Astron. J. 111, 535 (1996).

14. I. D. Karachentsev, S. N. Mitronova, V. E. Karachentseva,
Astron. Astrophys. 396, 431 (2002).

15. V. E. Karachentseva, Soobshch. Spets. Astron. Obs. Akad. Nauk SSSR 8,
3 (1973).

16. A. Kogut, C. Lineweaver, G.F. Smoot, et al., Astrophys. J. 419, 1 (1993).

 17. Yu. N. Kudrya, I. D. Karachentsev, V. E. Karachentseva,
and S. L. Parnovskii, Pis'ma Astron. Zh. 22, 330 (1996)
[Astron. Lett. 22, 295 (1996)].

18. Yu. N. Kudrya, I. D. Karachentsev, V. E. Karachentseva,
and S. L. Parnovskii, Pis'ma Astron. Zh. 23, 730 (1997)
[Astron. Lett. 23, 652 (1997)].

19. Yu. N. Kudrya, V. E. Karachentseva, I. D. Karachentsev, et al.,
Astron. Astrophys. 407, 889 (2003).

20. S. N. Mitronova, I. D. Karachentsev, V. E.
Karachentseva, et al., Bull. Spec. Astron. Obs. Russ. Acad. Sci. 57, 5 (2004).

 21. M. S. Roberts and M. P. Haynes, Ann. Rev. Astron.
Astrophys. 32, 115 (1994).

22. D. J. Schlegel, D. P. Finkbeiner, and M. Davis, Astrophys. J.
500, 525 (1998).

23. M. F. Skrutzkie, S. E. Schneider, R. Stiening, et al.,
in The Impact of Large Scale Near-IR Sky Surveys Ed. by F.
Garzon et al., (Kluwer, Dordrecht, 1997), Vol. 210, p. 25.

24. L. Stavely-Smith and R.D. Davies, Mon. Not. Roy. Astron.
Soc. 231, 833 (1988).

25. R. B. Tully and J. R. Fisher,
Astron. Astrophys. 54, 661 (1977).

 26. R. B. Tully and P. Fou\'que, Astrophys. J., Suppl. Ser. 58, 67 (1985).

27. A. V. Zasov, Astron. Zh. 51, 1225 (1974) [Sov. Astron. 18, 730 (1974)].

\newpage
\hoffset=-3cm
\renewcommand{\baselinestretch}{1.0}
\footnotesize
\begin{table}
\caption{Results of the HI observations of RFGC galaxies with
the 100-m radio telescope}

\begin{tabular}{rcrrrrrrrrr}
\hline
\multicolumn{1}{c}{RFGC}
&\multicolumn{1}{c}{RA(2000)DEC}
&\multicolumn{1}{c}{$a\times b$}
&\multicolumn{1}{c}{$B_t$}
&\multicolumn{1}{c}{$T$}
&\multicolumn{1}{c}{$F$}
&\multicolumn{1}{c}{$S_{max}$}
&\multicolumn{1}{c}{$V_h$}
&\multicolumn{1}{c}{$W_{50}$}
&\multicolumn{1}{c}{$W_{25}$}
&\multicolumn{1}{c}{$W_{20}$}\\ \hline
\multicolumn{1}{c}{1}&
\multicolumn{1}{c}{2}&
\multicolumn{1}{c}{3}&
\multicolumn{1}{c}{4}&
\multicolumn{1}{c}{5}&
\multicolumn{1}{c}{6}&
\multicolumn{1}{c}{7}&
\multicolumn{1}{c}{8}&
\multicolumn{1}{c}{9}&
\multicolumn{1}{c}{10}&
\multicolumn{1}{c}{11} \\ \hline

 161&004215.1$-$180940& 3.36$\times$ .30&14.5&7&19.1&139$\pm$ 4&1553 $\pm$2 &192&205&208 \\
 208&005332.6$+$025527& 1.71$\times$ .21&15.4&6& 4.2&   $\pm$ 4&4843 $\pm$3 &250&   &    \\
 280&011406.7$+$380723& 1.16$\times$ .16&16.0&5&    &   $\pm$ 2&            &   &   &    \\
 329&012737.1$+$510851& 0.87$\times$ .11&18.1&6& 2.1& 23$\pm$ 2&6096 $\pm$11&129&   &    \\
 411&015616.1$-$225404& 2.23$\times$ .27&14.9&3& 8.5& 68$\pm$ 9&1828 $\pm$1 &229&240&    \\
 458&020715.4$+$461520& 1.12$\times$ .13&16.4&7&    &   $\pm$ 5&            &   &   &    \\
 492&021732.5$-$113108& 1.22$\times$ .17&15.8&8& 3.6& 32$\pm$ 4&3957 $\pm$4 &210&   &    \\
 493&021735.9$-$214554& 0.95$\times$ .09&16.6&6& 4.1& 26$\pm$ 4&6423 $\pm$4 &264&   &    \\
 544&023052.8$+$432100& 1.27$\times$ .11&16.2&4&0.5:&   $\pm$ 3&6100:       &   &   &    \\
 620&025426.6$+$423900& 2.43$\times$ .15&15.4&7& 8.8& 68$\pm$ 6&2167 $\pm$2 &203&213&    \\
 668&030854.8$+$703349& 1.68$\times$ .24&17.0&3& 1.3& 37$\pm$ 9&4330 $\pm$5 &   &   &    \\
 734&032846.7$+$363323& 1.79$\times$ .20&16.5&4& 5.4& 41$\pm$ 6&4701 $\pm$4 &129&157&    \\
 860&043018.5$+$884616& 1.10$\times$ .13&16.3&6&    &   $\pm$ 4&            &   &   &    \\
 914&045213.2$-$182335& 1.53$\times$ .16&15.7&6& 1.4& 18$\pm$ 4&9573 $\pm$5 &415&   &    \\
 974&052320.9$+$854023& 0.83$\times$ .09&16.9&5&    &   $\pm$ 6&            &   &   &    \\
 982&052644.4$-$191236& 1.39$\times$ .19&15.8&5& 7.6& 30$\pm$ 6&8336 $\pm$3 &512&   &    \\
 987&053012.0$+$555216& 1.57$\times$ .21&18.0&9& 9.5& 60$\pm$ 3&2182 $\pm$4 &187&210&214 \\
1053&061021.8$+$504706& 1.38$\times$ .17&15.8&4& 3.3& 27$\pm$ 4&7775 $\pm$5 &420&   &    \\
1063&061318.7$+$530643& 1.19$\times$ .16&16.2&7&    &   $\pm$ 4&            &   &   &    \\
1064&061342.3$+$810425& 2.04$\times$ .24&15.6&4& 3.2& 28$\pm$ 4&4266 $\pm$7 &176&   &    \\
1119&063900.2$+$572258& 1.00$\times$ .12&16.4&5&    &   $\pm$ 5&            &   &   &    \\
1185&071554.5$+$675902& 2.02$\times$ .22&15.3&5& 4.7& 40$\pm$ 3&1153 $\pm$3 &172&183&    \\
1207&072225.0$+$491642& 0.84$\times$ .11&16.4&9&    &   $\pm$ 4&            &   &   &    \\
1280&075507.0$+$425728& 0.78$\times$ .09&16.8&6& 2.2& 15$\pm$ 4&7490 $\pm$5 &280&   &    \\
1282&075524.7$+$561002& 0.96$\times$ .11&16.7&5&    &   $\pm$ 4&            &   &   &    \\
1300&080009.6$+$562154& 2.33$\times$ .22&15.1&5& 5.0& 20$\pm$ 3&9095 $\pm$5 &629&   &    \\
1340&081359.5$+$454434& 5.17$\times$ .58&13.6&6&43.9&322$\pm$ 3& 559 $\pm$1 &170&189&194 \\
1351&081709.6$+$560249& 0.68$\times$ .09&16.9&6&    &   $\pm$ 4&            &   &   &    \\
1405&083812.0$+$455255& 0.87$\times$ .10&16.5&6& 1.7& 16$\pm$ 3&7070 $\pm$7 &264&   &    \\
1425&084421.6$+$093215& 1.23$\times$ .10&16.4&6& 4.0& 26$\pm$ 6&4064 $\pm$5 &226&   &    \\
1451&085419.0$+$542728& 0.90$\times$ .11&16.5&6& 2.1& 10$\pm$ 3&7449 $\pm$5 &350&   &    \\
1490&090738.9$+$283808& 1.10$\times$ .10&16.4&7& 1.0&  9$\pm$ 2&6682 $\pm$14&265&   &    \\
1524&091831.0$+$493244& 1.18$\times$ .15&16.0&6&    &   $\pm$11&            &   &   &    \\
1537&092145.1$+$641528& 4.31$\times$ .53&13.9&5&49.8&209$\pm$ 4&1574 $\pm$2 &308&336&341 \\
1543&092311.8$-$265631& 1.56$\times$ .17&15.8&5& 8: & 70$\pm$11&2410 $\pm$3 &193&   &    \\
1566&093140.8$-$160231& 1.84$\times$ .21&15.3&7&    &   $\pm$ 1&            &   &   &    \\
1583&093555.0$+$480848& 1.57$\times$ .22&15.6&4&    &   $\pm$13&            &   &   &    \\
\end{tabular}
\end{table}

\begin{table}
\begin{tabular}{rcrrrrrrrrr}
\hline
\multicolumn{1}{c}{1}&
\multicolumn{1}{c}{2}&
\multicolumn{1}{c}{3}&
\multicolumn{1}{c}{4}&
\multicolumn{1}{c}{5}&
\multicolumn{1}{c}{6}&
\multicolumn{1}{c}{7}&
\multicolumn{1}{c}{8}&
\multicolumn{1}{c}{9}&
\multicolumn{1}{c}{10}&
\multicolumn{1}{c}{11} \\ \hline

1612&094402.4$+$682212& 0.74$\times$ .09&16.8&6&    &   $\pm$ 5&           &   &   &    \\
1626&094650.4$+$794839& 2.49$\times$ .22&15.1&6&11.3& 86$\pm$ 3&1540$\pm$2 &170&183&186 \\
1711&100439.1$+$602759& 0.88$\times$ .10&16.6&6& 1.7& 16$\pm$ 2&2216$\pm$10&144&   &    \\
1735&101001.0$+$715221& 0.88$\times$ .10&16.6&6& 1.8& 19$\pm$ 3&6327$\pm$5 &189&   &    \\
1893&105124.0$-$195324& 4.26$\times$ .48&14.0&6&13.4& 86$\pm$13&2067$\pm$1 &229&   &    \\
1897&105236.0$+$394814& 0.68$\times$ .08&17.0&5&    &   $\pm$ 5&           &   &   &    \\
1940&110157.1$+$470540& 1.88$\times$ .15&15.8&5& 3.4& 19$\pm$ 3&6583$\pm$5 &378&   &    \\
1988&111327.8$+$072553& 0.73$\times$ .09&16.7&6&    &   $\pm$ 8&           &   &   &    \\
2056&113057.6$-$040507& 0.92$\times$ .10&16.7&7&    &   $\pm$ 5&           &   &   &    \\
2093&114227.4$+$513551& 3.70$\times$ .45&14.1&6& 9.3& 73$\pm$ 3& 975$\pm$2 &178&190&193 \\
2116&114836.0$+$434316& 1.87$\times$ .20&15.4&6& 2.8& 27$\pm$ 4& 735$\pm$5 &115&   &    \\
2258&122016.6$+$480811& 1.20$\times$ .17&15.8&9& 1.7&   $\pm$ 3& 839$\pm$4 &122&   &    \\
2296&122859.3$+$285143& 1.12$\times$ .15&16.1&5& 2.0& 17$\pm$ 3&8024$\pm$9 &435&   &    \\
2298&122908.6$+$575454& 1.34$\times$ .16&15.9&7& 5.2& 46$\pm$ 7& 799$\pm$4 &126&139&143 \\
2335&123621.1$+$255906&15.90$\times$1.85&10.6&4&161 &420$\pm$ 5&1226$\pm$1 &506&521&524 \\
2354&124128.8$-$031514& 0.67$\times$ .09&17.1&8& 1.3& 15$\pm$ 3&1803$\pm$6 &125&   &    \\
2390&125105.0$-$082623& 1.12$\times$ .16&16.1&6&    &   $\pm$ 6&           &   &   &    \\
2437&130106.7$-$032235& 0.83$\times$ .11&16.8&7& 2.4& 23$\pm$ 5&2900$\pm$6 &156&175&    \\
2443&130207.9$+$584159& 3.92$\times$ .32&14.5&6&    &   $\pm$ 3&           &   &   &    \\
2460&130514.9$-$002230& 1.09$\times$ .13&16.3&7&    &   $\pm$ 3&           &   &   &    \\
2463&130548.0$+$462743& 1.37$\times$ .19&15.7&4&    &   $\pm$ 3&           &   &   &    \\
2553&132409.6$-$175407& 1.25$\times$ .10&16.4&7& 1.9& 26$\pm$ 6&6994$\pm$15&200&   &    \\
2559&132525.7$+$044600& 1.01$\times$ .10&16.7&5&    &   $\pm$ 8&           &   &   &    \\
2574&132820.2$-$114703& 1.85$\times$ .24&15.4&4&    &   $\pm$ 3&           &   &   &    \\
2592&133341.5$-$111418& 0.99$\times$ .07&17.0&7&    &   $\pm$ 4&           &   &   &    \\
2626&134043.3$-$173314& 0.65$\times$ .07&17.2&8&    &   $\pm$ 7&           &   &   &    \\
2707&140425.4$-$002725& 0.90$\times$ .10&16.6&7& 1.5& 13$\pm$ 2&7380$\pm$6 &301&   &    \\
2713&140621.6$-$054313& 1.28$\times$ .17&15.8&4&    &   $\pm$ 4&           &   &   &    \\
2737&141312.0$-$072648& 0.67$\times$ .08&17.3&9&    &   $\pm$ 6&           &   &   &    \\
2754&141741.0$-$052748& 1.29$\times$ .17&15.9&5& 6.1& 36$\pm$ 9&7057$\pm$5 &308&   &    \\
2757&141833.6$-$050913& 1.15$\times$ .10&16.4&7& 3.2& 22$\pm$ 5&7010$\pm$11&229&   &    \\
2786&143007.2$-$062543& 1.46$\times$ .17&15.9&4&    &   $\pm$ 4&           &   &   &    \\
2856&144457.6$+$405234& 1.12$\times$ .16&15.9&6& 1.8& 14$\pm$ 3&5640$\pm$12&216&   &    \\
2902&150032.6$+$491027& 0.76$\times$ .10&16.7&4&    &   $\pm$ 4&           &   &   &    \\
2915&150438.9$-$181337& 0.84$\times$ .09&16.9&7&    &   $\pm$ 7&           &   &   &    \\
2943&151543.2$-$002524& 0.67$\times$ .09&17.0&7&    &   $\pm$ 5&           &   &   &    \\
2944&151543.2$-$102750& 0.83$\times$ .09&16.8&7&    &   $\pm$ 6&           &   &   &    \\

\end{tabular}
\end{table}

\begin{table}
\begin{tabular}{rcrrrrrrrrr}
\hline
\multicolumn{1}{c}{1}&
\multicolumn{1}{c}{2}&
\multicolumn{1}{c}{3}&
\multicolumn{1}{c}{4}&
\multicolumn{1}{c}{5}&
\multicolumn{1}{c}{6}&
\multicolumn{1}{c}{7}&
\multicolumn{1}{c}{8}&
\multicolumn{1}{c}{9}&
\multicolumn{1}{c}{10}&
\multicolumn{1}{c}{11} \\ \hline
2955&151836.7$-$011101&1.04$\times$ .11&16.8&6&   &  $\pm$ 6&            &   &   &    \\
2978&152628.8$+$411731&2.26$\times$ .24&15.1&5&5.6&37$\pm$ 5&2481 $\pm$1 &189&   &    \\
3001&153443.0$+$082003&1.48$\times$ .12&16.3&6&2.6&20$\pm$ 3&5803 $\pm$4 &278&   &    \\
3007&153654.0$+$511720&1.10$\times$ .11&16.4&5&4.2&30$\pm$ 4&5847 $\pm$5 &209&   &    \\
3154&163231.2$+$674449&1.12$\times$ .12&16.4&4&4.0&21$\pm$ 3&3423 $\pm$4 &255&   &    \\
3202&165140.3$+$532422&0.68$\times$ .09&16.8&8&   &  $\pm$ 4&            &   &   &    \\
3210&165450.4$+$702617&1.37$\times$ .19&15.7&5&   &  $\pm$ 3&            &   &   &    \\
3228&170247.8$+$444731&0.65$\times$ .08&17.3&8&   &  $\pm$ 4&            &   &   &    \\
3230&170338.9$+$454832&0.83$\times$ .10&16.9&7&   &  $\pm$ 5&            &   &   &    \\
3248&171145.3$-$254451&0.98$\times$ .10&16.7&5&3.4&17$\pm$ 4&6330 $\pm$6 &353&   &    \\
3249&171149.2$+$473937&0.85$\times$ .10&16.7&6&2.5&15$\pm$ 2&8320 $\pm$15&248&   &    \\
3297&173721.6$+$602537&1.12$\times$ .11&16.4&5&2.5&22$\pm$12&6220 $\pm$6 &254&   &    \\
3310&174810.8$+$530908&0.95$\times$ .10&16.5&6&1.5&20$\pm$ 3&6744 $\pm$9 &241&   &    \\
3333&180645.1$+$874835&1.77$\times$ .21&15.6&4&   &  $\pm$ 8&            &   &   &    \\
3377&183312.7$+$525655&1.34$\times$ .11&16.3&4&4.7&25$\pm$ 6&8418:       &   &   &    \\
3680&210236.0$-$134753&1.68$\times$ .20&15.6&3&   &  $\pm$ 2&            &   &   &    \\
3870&220604.3$-$261107&1.27$\times$ .13&16.1&6&   &  $\pm$ 2&            &   &   &    \\
4160&233900.0$+$493531&0.83$\times$ .10&16.7&6&1.2&14$\pm$ 3&9319 $\pm$10&   &   &    \\
4181&234438.4$-$273936&1.14$\times$ .16&16.1&5&   &  $\pm$ 4&            &   &   &    \\
4214&235321.6$+$860141&0.96$\times$ .09&16.7&6&2.2&12$\pm$ 3&5739 $\pm$10&361&   &    \\

\hline
\end{tabular}
\end{table}

\newpage
\hoffset=-3cm
\renewcommand{\baselinestretch}{1.0}
\footnotesize
\begin{table}
\caption{Results of the HI observations of 2MFGC galaxies with the
100-m radio telescope}
\begin{tabular}{lcrrrrrrrrr}
\hline
\multicolumn{1}{c}{Name}
&\multicolumn{1}{c}{RA(2000)DEC}
&\multicolumn{1}{c}{$a\times b$}
&\multicolumn{1}{c}{$B_t$}
&\multicolumn{1}{c}{$T$}
&\multicolumn{1}{c}{$F$}
&\multicolumn{1}{c}{$S_{max}$}
&\multicolumn{1}{c}{$V_h$}
&\multicolumn{1}{c}{$W_{50}$}
&\multicolumn{1}{c}{$W_{25}$}
&\multicolumn{1}{c}{$W_{20}$}\\ \hline
\multicolumn{1}{c}{1}&
\multicolumn{1}{c}{2}&
\multicolumn{1}{c}{3}&
\multicolumn{1}{c}{4}&
\multicolumn{1}{c}{5}&
\multicolumn{1}{c}{6}&
\multicolumn{1}{c}{7}&
\multicolumn{1}{c}{8}&
\multicolumn{1}{c}{9}&
\multicolumn{1}{c}{10}&
\multicolumn{1}{c}{11} \\ \hline

UGC256       &002656.6$+$500150.7&1.55$\times$ .29&15.28&4& 1.5&11$\pm$3&          &   &   & \\
UGC1928      &022811.2$+$435109.7&1.36$\times$ .30&16.47&5& 2.6&16$\pm$3&6561$\pm$7&432&   & \\
UGC69A       &032702.2$+$725038.3&2.10$\times$ .49&     &3&17.7&83$\pm$4&2074$\pm$3&243&257&262\\
UGC2736      &032627.6$+$403028.5&1.57$\times$ .31&14.64&2&    &  $\pm$5&          &   &   & \\
MCG-03-13-016&045013.7$-$171557.8&1.83$\times$ .53&14.25&4& 5.2&22$\pm$3&9248$\pm$8&608&   & \\
UGC3127      &044025.9$-$020112.6&2.10$\times$ .42&14.94&7&11.0&74$\pm$5&3345$\pm$4&192&205&209\\
UGC3342      &054429.7$+$691756.3&1.74$\times$ .39&14.99&5& 4.4&31$\pm$4&3890$\pm$4&294&   & \\
IRAS05442+46 &054809.6$+$461531.0&0.77$\times$ .60&16.23&5& 3.2&27$\pm$4&6083$\pm$7&190&   & \\
UGC4258      &081047.6$+$465445.7&1.43$\times$ .38&18.16&6& 5.3&37$\pm$5&3124$\pm$4&208&222& \\
UGC6390      &112242.0$+$640358.7&2.07$\times$ .30&19.22&6& 9.3&91$\pm$4& 982$\pm$3&163&177&180\\
UGC6575      &113626.5$+$581122.0&1.88$\times$ .44&14.57&5&14.6&89$\pm$3&1216$\pm$5&206&222& \\
UGC6894      &115524.4$+$543926.3&1.45$\times$ .25&15.28&5& 4.0&39$\pm$5& 850$\pm$4&128&   & \\
NGC6244      &164803.9$+$621201.6&1.54$\times$ .32&14.45&1&    &  $\pm$4&          &   &   & \\
UGC10713     &170433.9$+$722647.5&1.83$\times$ .32&13.97&3&12.7&89$\pm$4&1072$\pm$2&230&242&244\\

\hline
\end{tabular}
\end{table}

\small
\begin{table}
\caption{Optical and HI parameters of galaxies}
\begin{tabular}{rrrrrrrrr}  \hline
\multicolumn{1}{r}{RFGC}
&\multicolumn{1}{c}{$V_{LG}$}
&\multicolumn{1}{r}{$W_c$}
&\multicolumn{1}{c}{$log(M_{HI}/M_{\sun})$}
&\multicolumn{1}{r}{$log(M_{25}/M_{\sun})$}
&\multicolumn{1}{r}{$log(L/L_{\sun})$}
&\multicolumn{1}{r}{$M_{HI}/L$}
&\multicolumn{1}{r}{$M_{25}/L$}
&\multicolumn{1}{r}{$M_{HI}/M_{25}$} \\
\hline
\multicolumn{1}{c}{1}&
\multicolumn{1}{c}{2}&
\multicolumn{1}{c}{3}&
\multicolumn{1}{c}{4}&
\multicolumn{1}{c}{5}&
\multicolumn{1}{c}{6}&
\multicolumn{1}{c}{7}&
\multicolumn{1}{c}{8}&
\multicolumn{1}{c}{9}\\ \hline
    &     &    &      &      &      &     &     &     \\
 161&1623 & 191&  9.3& 10.1&  9.6&  0.57&  3.76&  0.15 \\
 208&4998 & 246&  9.6& 10.6& 10.1&  0.33&  2.89&  0.11 \\
 329&6352 & 126&  9.5&  9.9& 10.3&  0.20&  0.48&  0.42 \\
 411&1824 & 228&  9.1& 10.2&  9.4&  0.43&  5.68&  0.08 \\
 492&3985 & 207&  9.4& 10.2&  9.8&  0.42&  2.83&  0.15 \\
 493&6407 & 258&  9.8& 10.5&  9.9&  0.87&  3.55&  0.25 \\
 620&2357 & 202&  9.3& 10.2&  9.7&  0.36&  2.89&  0.13 \\
 734&4850 & 127&  9.7& 10.1& 10.5&  0.16&  0.39&  0.41 \\
 914&9438 & 402&  9.7& 11.2& 10.6&  0.12&  4.06&  0.03 \\
 982&8170 & 498& 10.3& 11.3& 10.4&  0.79&  8.23&  0.10 \\
 987&2321 & 186&  9.3& 10.2& 10.3&  0.11&  0.79&  0.14 \\
1053&7874 & 409&  9.9& 11.2& 10.7&  0.19&  3.48&  0.05 \\
1064&4476 & 174&  9.4& 10.3& 10.2&  0.16&  1.31&  0.12 \\
1185&1305 & 171&  8.5&  9.8&  9.1&  0.29&  5.00&  0.06 \\
1280&7509 & 273&  9.7& 10.5& 10.0&  0.56&  3.56&  0.16 \\
1300&9182 & 610& 10.2& 11.8& 10.9&  0.25&  8.02&  0.03 \\
1340& 588 & 170&  8.8&  9.8&  9.0&  0.57&  5.85&  0.10 \\
1405&7095 & 258&  9.6& 10.5& 10.0&  0.35&  2.96&  0.12 \\
1425&3898 & 223&  9.4& 10.2&  9.7&  0.55&  3.65&  0.15 \\
1451&7517 & 342&  9.7& 10.8& 10.0&  0.45&  5.35&  0.08 \\
1490&6608 & 259&  9.3& 10.5& 10.0&  0.17&  3.17&  0.05 \\
1537&1692 & 306&  9.8& 10.7&  9.8&  0.85&  7.54&  0.11 \\
1543&2114 & 191&  9.2& 10.0&  9.4&  0.62&  3.95&  0.16 \\
1626&1732 & 169&  9.2&  9.9&  9.4&  0.58&  3.50&  0.16 \\
1711&2314 & 143&  8.6&  9.5&  9.0&  0.40&  3.22&  0.13 \\
1735&6483 & 185&  9.5& 10.2&  9.9&  0.36&  1.69&  0.21 \\
1893&1796 & 227&  9.3& 10.5&  9.8&  0.26&  4.23&  0.06 \\
1940&6617 & 370&  9.8& 11.1& 10.3&  0.32&  5.97&  0.05 \\
2093&1043 & 177&  8.6& 10.0&  9.3&  0.23&  4.81&  0.05 \\
2116& 765 & 115&  7.8&  9.1&  8.5&  0.21&  4.14&  0.05 \\
2258& 903 & 122&  7.8&  9.1&  8.4&  0.22&  4.53&  0.05 \\
2296&7997 & 424&  9.7& 11.1& 10.2&  0.31&  7.09&  0.04 \\
2298& 912 & 126&  8.3&  9.2&  8.4&  0.67&  5.21&  0.13 \\
2335&1189 & 504& 10.0& 11.5& 10.8&  0.15&  5.65&  0.03 \\
2354&1639 & 124&  8.2&  9.1&  8.5&  0.50&  4.42&  0.11 \\
2437&2749 & 155&  8.9&  9.6&  9.0&  0.69&  3.79&  0.18 \\
2553&6807 & 195&  9.6& 10.4& 10.2&  0.23&  1.47&  0.15 \\
\end{tabular}
\end{table}

\begin{table}
\begin{tabular}{rrrrrrrrr}  \hline
\multicolumn{1}{r}{RFGC}
&\multicolumn{1}{c}{$V_{LG}$}
&\multicolumn{1}{r}{$W_c$}
&\multicolumn{1}{c}{$log(M_{HI}/M_{\sun})$}
&\multicolumn{1}{r}{$log(M_{25}/M_{\sun})$}
&\multicolumn{1}{r}{$log(L/L_{\sun})$}
&\multicolumn{1}{r}{$M_{HI}/L$}
&\multicolumn{1}{r}{$M_{25}/L$}
&\multicolumn{1}{r}{$M_{HI}/M_{25}$} \\ \hline
\multicolumn{1}{c}{1}&
\multicolumn{1}{c}{2}&
\multicolumn{1}{c}{3}&
\multicolumn{1}{c}{4}&
\multicolumn{1}{c}{5}&
\multicolumn{1}{c}{6}&
\multicolumn{1}{c}{7}&
\multicolumn{1}{c}{8}&
\multicolumn{1}{c}{9}\\ \hline
2707&7290 & 294&  9.5& 10.6& 10.0&  0.30&  3.85&  0.08 \\
2754&6959 & 301& 10.1& 10.8& 10.2&  0.74&  3.73&  0.20 \\
2757&6914 & 224&  9.8& 10.4& 10.1&  0.49&  2.18&  0.23 \\
2856&5753 & 212&  9.4& 10.3& 10.0&  0.25&  2.21&  0.11 \\
2978&2625 & 187&  9.2& 10.2&  9.7&  0.31&  2.88&  0.11 \\
3001&5831 & 273&  9.6& 10.7& 10.0&  0.35&  4.24&  0.08 \\
3007&6027 & 205&  9.8& 10.3&  9.9&  0.77&  2.41&  0.32 \\
3154&3657 & 252&  9.3& 10.3&  9.5&  0.70&  5.96&  0.12 \\
3248&6302 & 346&  9.7& 10.9& 10.7&  0.11&  1.77&  0.06 \\
3249&8548 & 241&  9.9& 10.5& 10.1&  0.59&  2.47&  0.24 \\
3297&6474 & 249&  9.6& 10.5& 10.0&  0.39&  2.94&  0.13 \\
3310&6998 & 236&  9.5& 10.4& 10.0&  0.29&  2.52&  0.11 \\
4214&5985 & 354&  9.5& 10.8& 10.1&  0.26&  4.70&  0.06 \\

\hline
\end{tabular}
\end{table}

\small
\begin{table*}
\caption{Characteristics of the distributions of flat
galaxies in integrated parameters}
\begin{tabular}{lrrrr} \\ \hline
 Parameter             &   Mean    &Standard&  Skewness & Kurtosis  \\
		      &              &deviation &             &          \\
\hline
 log$(W_{50}$)   &       2.36   &   0.17   &   0.33   &    $-$0.40    \\
 log$(A_{25}$)   &       1.17   &   0.29   &   $-$0.79   &    0.50    \\
 log$(L_B)$     &      9.79   &   0.59   &   $-$0.65   &    0.01    \\
 log$(M_{HI})$    &       9.38   &   0.56   &   $-$1.10   &    0.94    \\
 log($M_{25}$)   &      10.35   &   0.59   &   $-$0.11   &    0.10     \\
 log($M_{HI}/L_B$) &      $-$0.42   &   0.23   &   $-$0.16   &    $-$0.77     \\
 log($M_{HI}/A_{25}^2$)&   7.03   &   0.23   &    $-$0.20   &    $-$0.82     \\
 log($M_{25}/L_B)$ &       0.56   &   0.21   &   $-$0.43   &    0.07     \\
 log($M_{HI}/M_{25}$)&      $-$0.98   &   0.28   &   $-$0.07   &    $-$0.42     \\
 log($A_{25}\cdot W_{50}$)&      3.53   &   0.43   &   $-$0.37   &    0.31     \\
\hline
\end{tabular}
\end{table*}

\begin{table}[h]
\caption{Distribution of the mean parameters in morphological type for RFGC galaxies}
\begin{tabular}{rlcccc} \\ \hline
N & Type & log$(M_{HI}/A_{25}^2$) & log($M_{25}/L_B)$ & log($M_{HI}/M_{25})$&
log($M_{HI}/L_B$)\\
\hline
18 & Sb,bc,c & 7.01$\pm$ 0.06& 0.60$\pm$ 0.06 &$-$1.03$\pm$ 0.08& $-$0.43$\pm$0.06  \\
20 & Scd     & 7.03$\pm$ 0.06& 0.53$\pm$ 0.03 &$-$0.96$\pm$ 0.06& $-$0.43$\pm$0.05  \\
12 & Sd,dm,m & 7.07$\pm$ 0.05& 0.55$\pm$ 0.05 &$-$0.93$\pm$ 0.06& $-$0.38$\pm$0.07  \\
\hline
\end{tabular}
\end{table}

\begin{table*}[h]
\caption{Parameters of the linear regression $y = kx + c$ for the global
characteristics of RFGC galaxies}
\begin{tabular}{rllrrrr} \\ \hline
$N$ &
$y$ &
$x$ &
$\rho(xy)$ &
$\sigma(y)$ &
$k\pm \sigma_k$ &
$c\pm\sigma_c$\\ \hline
1& log$(M_{HI})$  & log$(A_{25})$   &  0.92   & 0.22  &   1.80$\pm$ 0.11&   7.26$\pm$ 0.13  \\
2& log$(M_{25})$       & log$(A_{25})$   &  0.93   & 0.22  &   1.92$\pm$ 0.11&   8.10$\pm$ 0.13  \\
3& log$(L_B)$         & log$(A_{25})$   &  0.97   & 0.15  &   2.00$\pm$ 0.08&   7.45$\pm$ 0.09  \\
4& log$(M_{HI})$       & log$(L_B)$     &  0.92   & 0.22  &   0.87$\pm$ 0.05&   0.81$\pm$ 0.52   \\
5& log$(M_{25})$       & log$(L_B)$     &  0.94   & 0.20  &   0.94$\pm$ 0.05&   1.14$\pm$ 0.48   \\
6& log$(A_{25})$       & log$(W_{50})$   &  0.77   & 0.19  &   1.28$\pm$ 0.15&  $-$1.86$\pm$ 0.37   \\
7& log$(M_{25}/L_B)$     & log$(W_{50}) $  &  0.39   & 0.19  &   0.47$\pm$ 0.16&  $-$0.54$\pm$ 0.38  \\
8& log$(M_{HI}/L_B)$     & log$(W_{50}) $  & $-$0.24   & 0.23  &  $-$0.32$\pm$ 0.19&  0.34$\pm$ 0.45  \\
9& log$(M_{25}/M_{HI})$   & log$(W_{50})$   &  $-$0.49   & 0.24  &   $-$0.79$\pm$ 0.20&  0.88$\pm$ 0.48  \\
10& log$(M_{HI}/A_{25}^2$)& log$(W_{50})$   & $-$0.05   & 0.23  &  $-$0.07$\pm$ 0.19&   7.19$\pm$ 0.45  \\
11& log$(M_{HI})$       & log($A_{25}\cdot W_{50})$& 0.91   & 0.23  &   1.19$\pm$ 0.08&   5.19$\pm$ 0.27  \\
12& log$(M_{HI}) $      & log($W_{50})$   &  0.76   & 0.37  &   2.49$\pm$ 0.31&   3.48$\pm$ 0.73  \\
13& log$(M_{25}/L_B)$ & log$(V_{LG})$ &  $-$0.32 & 0.20  & $-0.20\pm 0.08$&  1.25$\pm$ 0.30  \\
\hline
\end{tabular}
\end{table*}

\newpage

\begin{figure*}[hbt]
 \vbox{\includegraphics{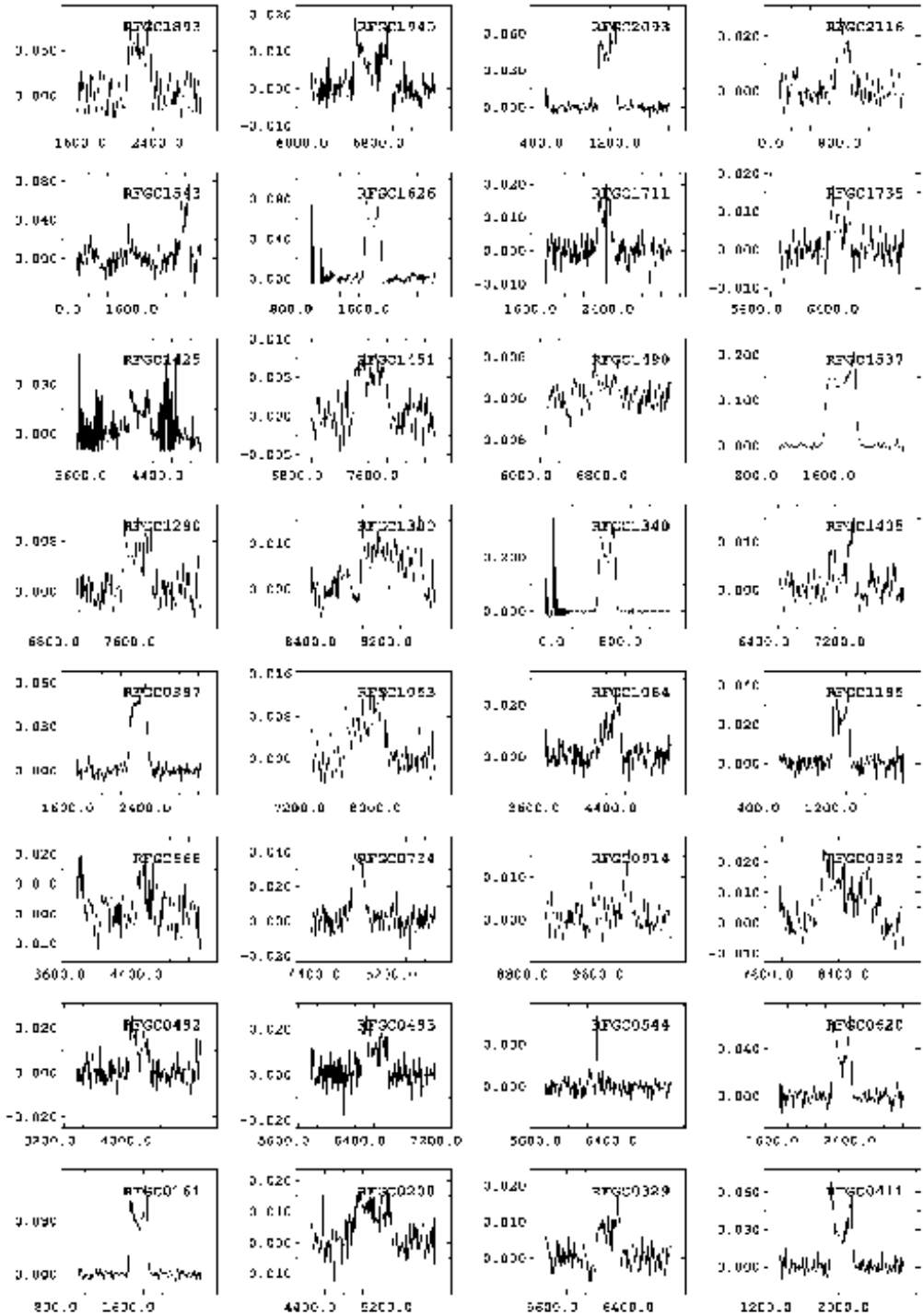}}
\vspace{20.0cm}
\caption{HI profiles for 51 RFGC galaxies from Table 1}
\end{figure*}

\newpage

\setcounter{figure}{0}
\begin{figure*}[hbt]
 \vbox{\includegraphics{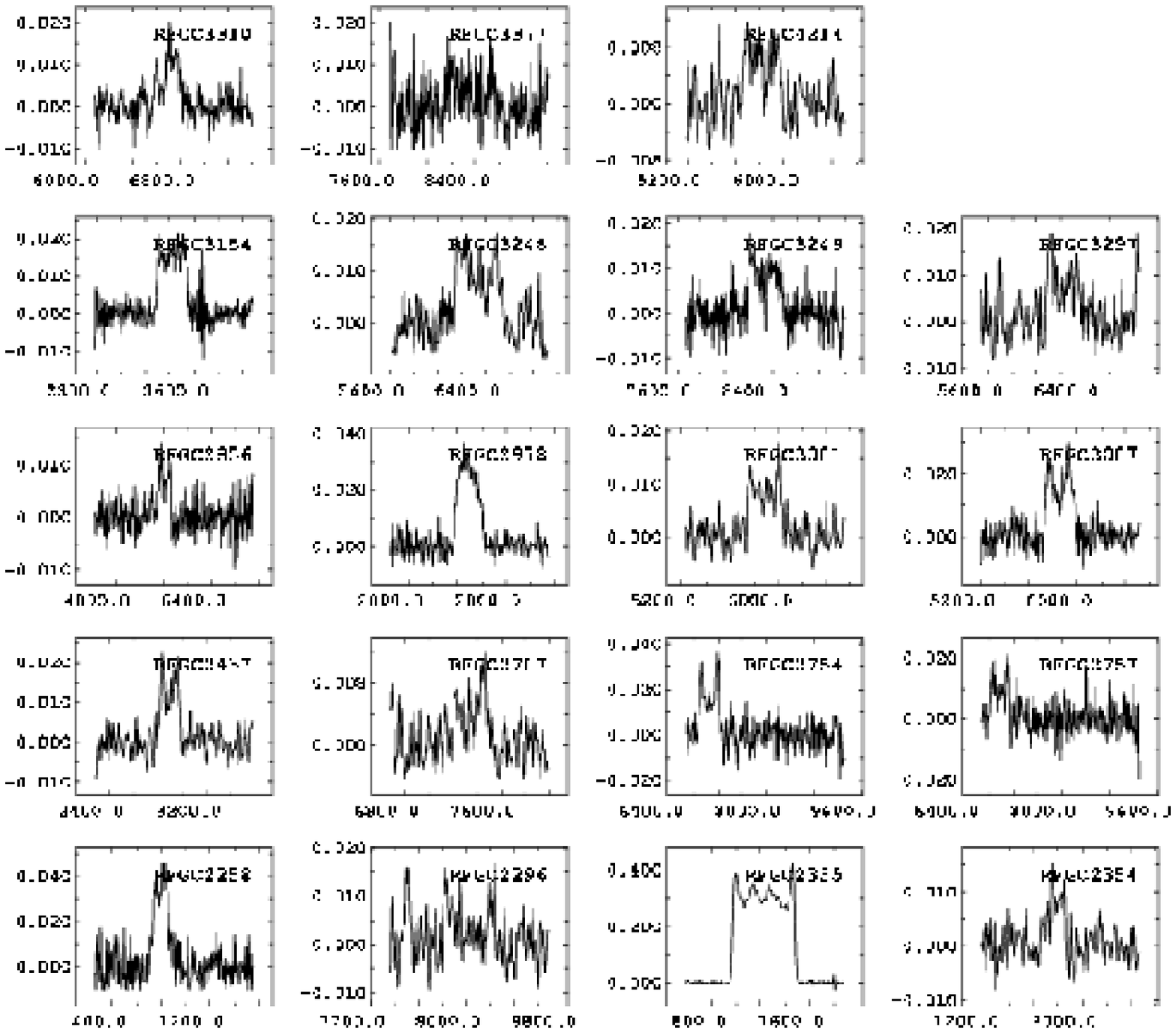}}
\vspace{10.7cm}
\caption{Contd.}
\end{figure*}

\begin{figure*}[hbt]
 \vbox{\includegraphics{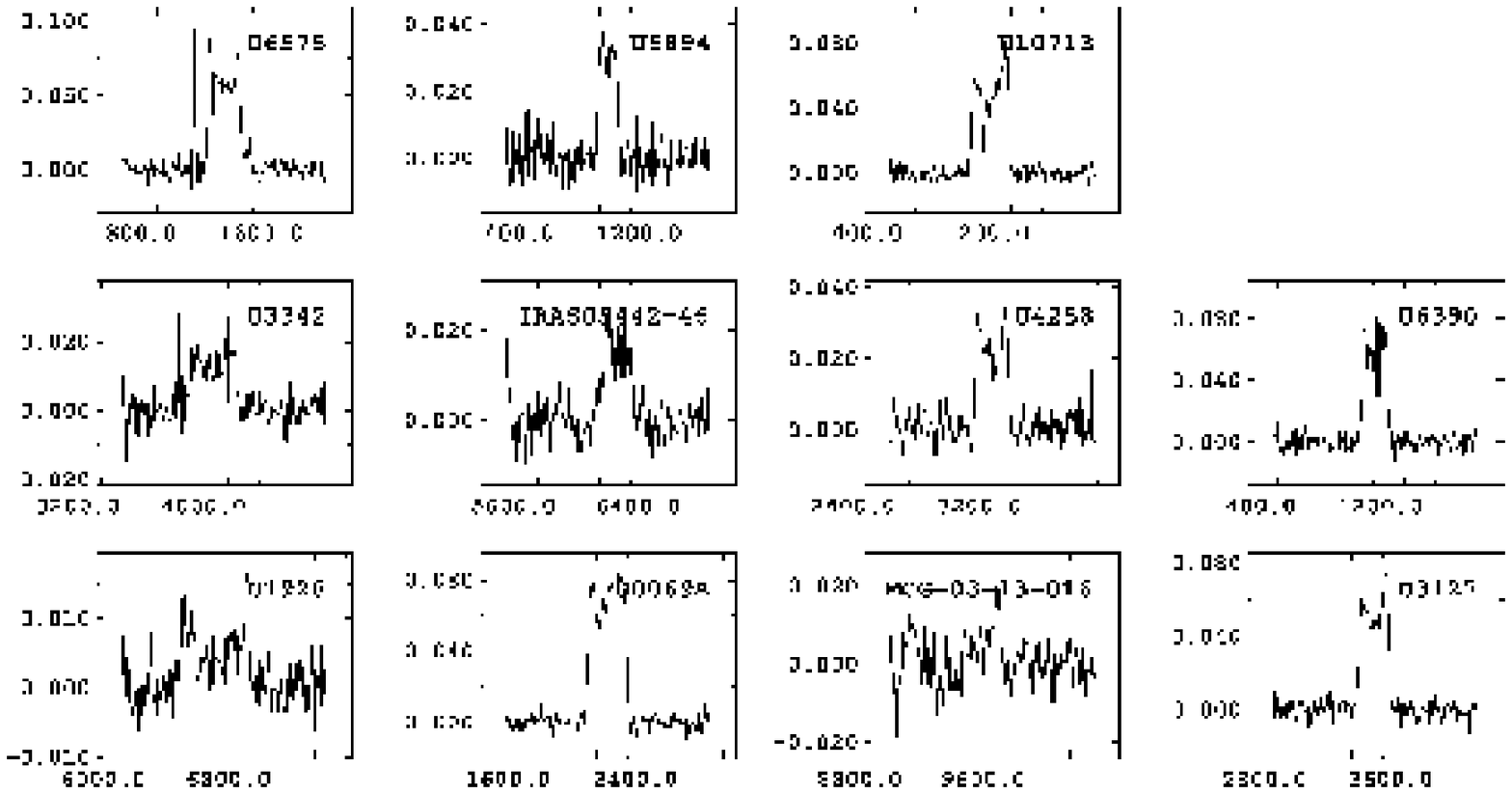}}
\vspace{8.0cm}
\caption{HI profiles for 11 2MFGC galaxies from Table 2}
\end{figure*}

\begin{figure*}[hbt]
 \vbox{\includegraphics{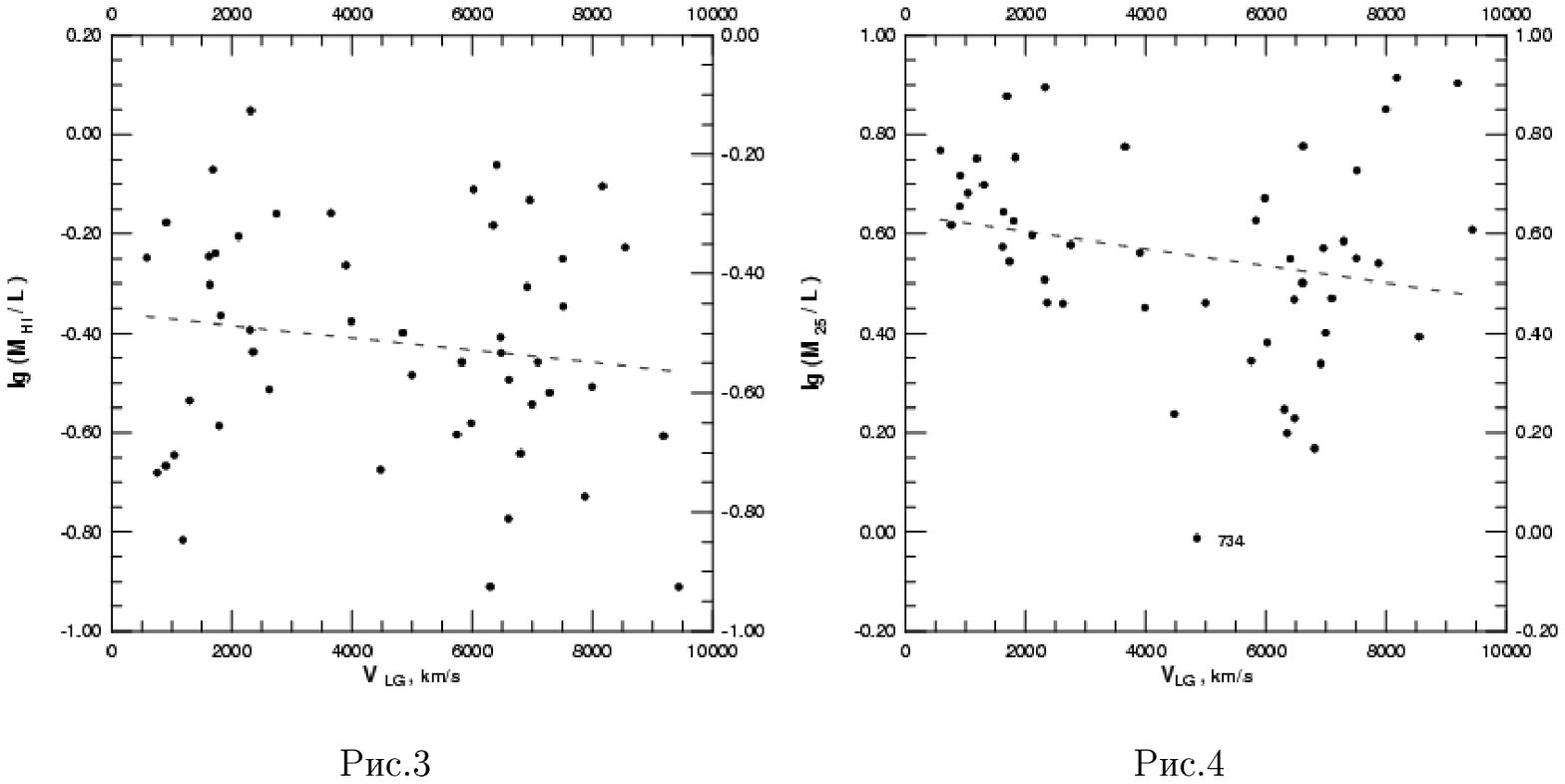}}
\vspace{6.0cm}
\caption{HI  mass-to-luminosity ratio (in solar units)
versus radial velocity (in km s$^{-1}$ )}

\caption{Total mass-to-luminosity ratio (in solar units)
versus radial velocity (in km s$^{-1}$ )}
\end{figure*}

\begin{figure*}
 \vbox{\includegraphics{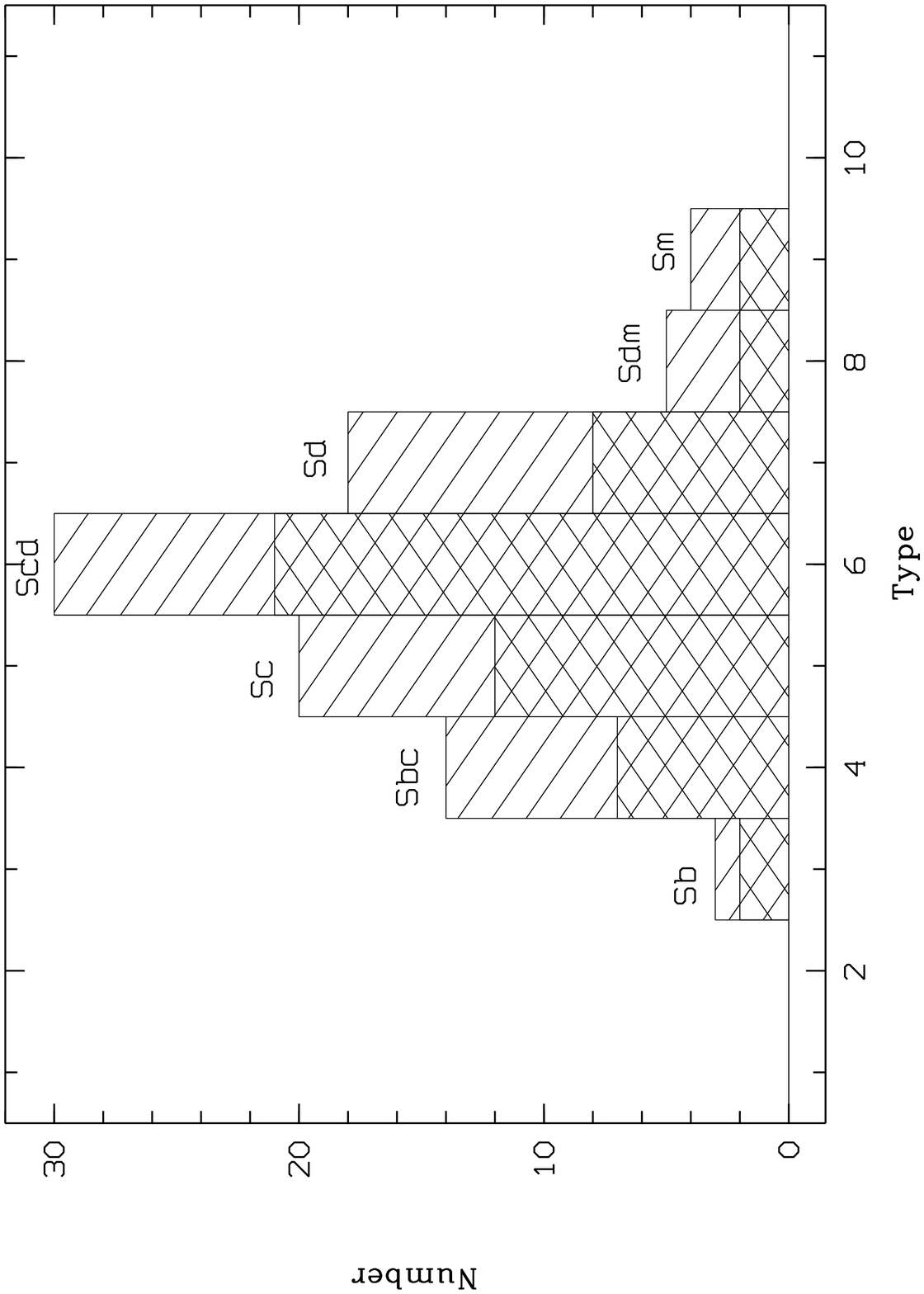}}
\vspace{5.0cm}
\caption{Distribution of galaxies in morphological type}
\end{figure*}

\newpage
\begin{figure*}[hbt]
 \vbox{\includegraphics{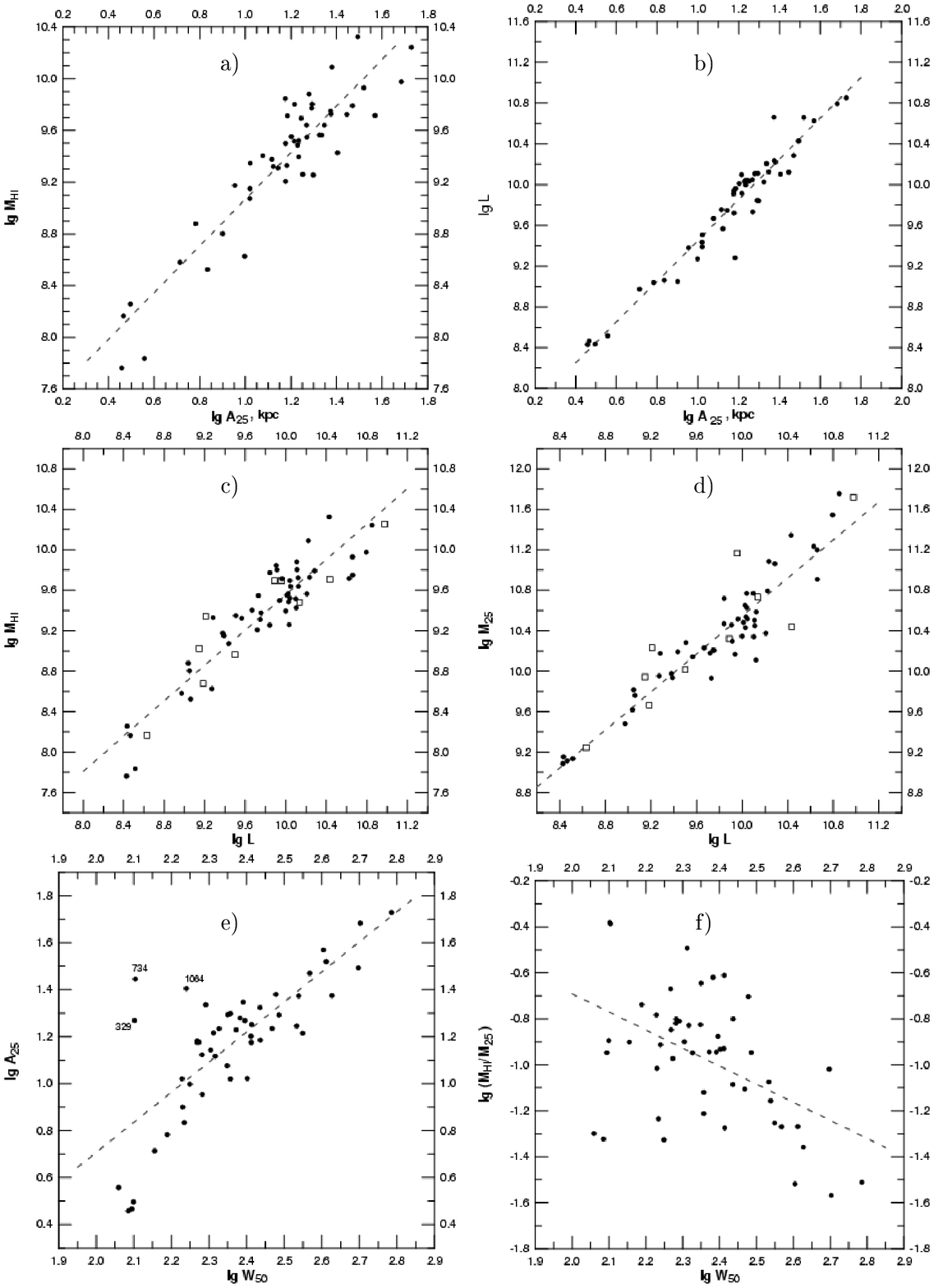}}
\vspace{16.5cm}
\caption{Bivariate relations between the global parameters reduced
to solar units: (a) logarithm of the HI mass versus linear diameter
(in kpc); (b) luminosity
versus linear diameter; (c) HI mass versus luminosity; (d)
indicative mass versus luminosity, the squares mark the 2MFGC
galaxies; (e) linear diameter of the galaxy versus HI line width
(in km s$^{-1}$ ); and (f) HI mass fraction in the galaxy versus
HI line width. The dashes represent the regression lines.}
\end{figure*}

\begin{figure*}[hbt]
 \vbox{\includegraphics{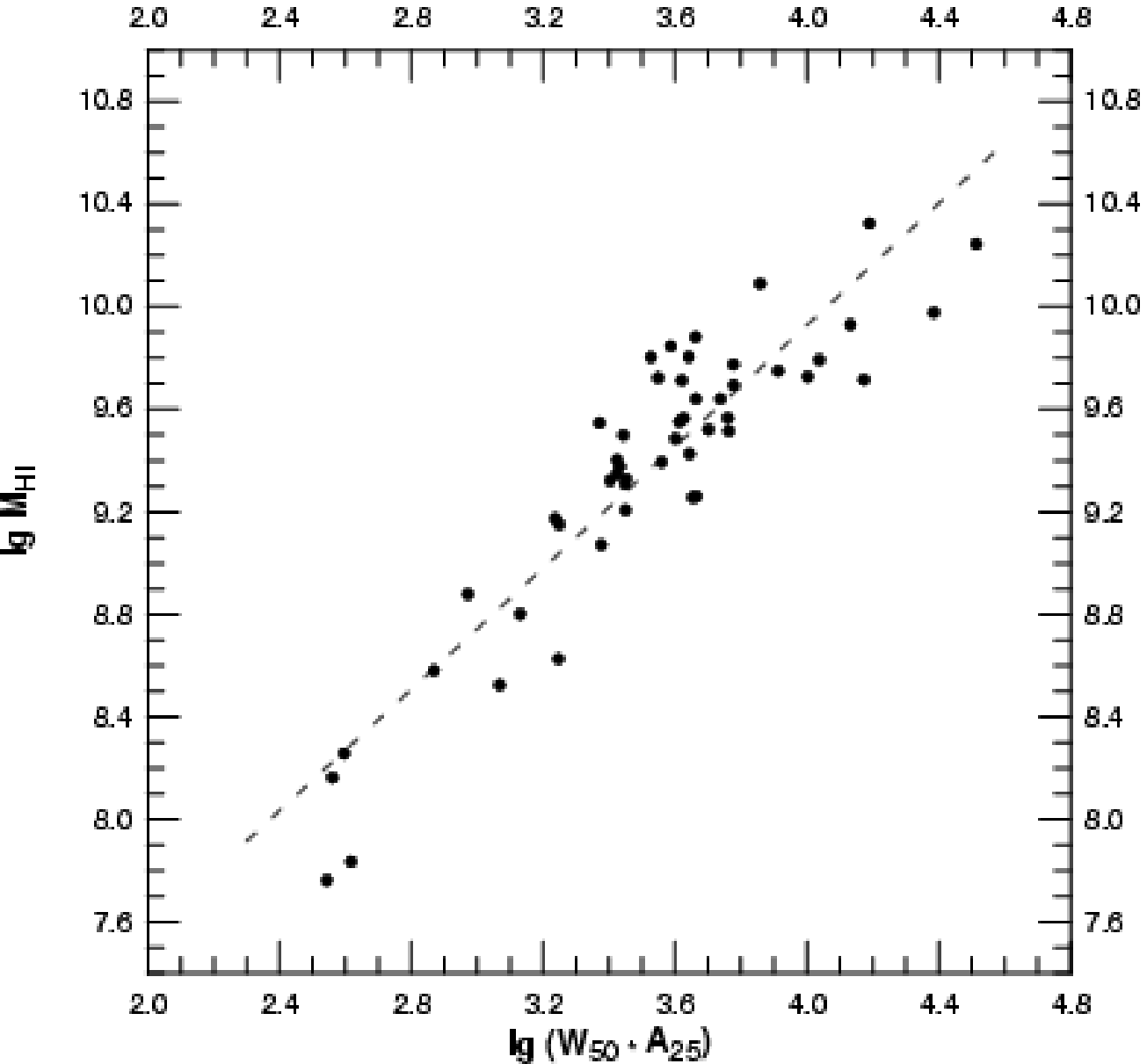}}
\vspace{5.3cm}
\caption{
HI mass versus specific angular momentum}
\end{figure*}

\end{document}